\documentclass{ws-ijmpe}
\usepackage[super,compress]{cite}
\usepackage{xcolor}
\usepackage{amsmath}
\usepackage{listings}
\usepackage{soul}
\usepackage{lipsum} 
\begin{document}
\large 

\markboth{ Ashwini Kumar, 
      Bushra Ali, Gauri Devi, B. K. Singh, Shakeel Ahmad}{Pseudorapidity Dependence of Multiplicity Fluctuations in High-Energy Collisions with System Size and Beam Energies}

\catchline{}{}{}{}{}

\title{Pseudorapidity Window Size Dependence of Multiplicity Fluctuations in High-Energy Collisions with System Size and Beam Energies}

\author{Ashwini Kumar\footnote{ashwini.physics@gmail.com}}
\address{Department of Physics and Electronics, Dr. Rammanohar Lohia Avadh University\\
Ayodhya, 224001, U.P., INDIA\\
}
\author{Gauri Devi, B. K. Singh}
\address{Department of Physics, Banaras Hindu University, Varanasi-221005, U.P., INDIA}
\author{Bushra Ali, Shakeel Ahmad\footnote{Shakeel.ahmad@cern.ch}}
\address{Department of Physics, Aligarh Muslim University,  \\
Aligarh-202002, U.P. INDIA\\
}

\maketitle

\begin{history}
\received{Day Month Year}
\revised{Day Month Year}
\end{history}

\begin{abstract}
\large  
  An investigation of the critical behavior of strongly interacting QCD matter has been performed by analyzing fluctuation observables on event-by-event (ebe) basis measured in high-energy collision experiments. The fluctuation analysis is performed  using nuclear interactions at different target sizes and at different colliding beam energies as a function of varying width of pseudorapidity interval. For the sake of comparison, event-by-event multiplicity fluctuations in hadronic and heavy-ion collisions (p-H, p-A and A-B) are studied within the framework of the Lund Monte Carlo based FRITIOF model. Charged particle multiplicity and the variance of the multiplicity distribution are estimated for the interactions involving different target sizes and beam momenta i.e., p-H, p-CNO, p-AgBr at 200A GeV/c and $^{16}$O-AgBr collisions at 14.6, 60 and 200A GeV/c. Further, multiplicity fluctuations are quantified in terms of intensive quantity, the scaled variances $\omega$ and the strongly intensive quantity $\Sigma_{FB}$ derived from the charged particle multiplicity and the width of the multiplicity distribution. Strongly intensive quantity $\Sigma_{FB}$ is a quantity of great significance to extract information about short and long-range multiplicity correlations. Furthermore, the collision centrality and centrality bin width dependent behavior of the multiplicity fluctuation have been examined in the framework of Lund Monte Carlo based FRITIOF model. Results based on the fluctuation analysis carried out in the present study are interpreted in terms of dynamics of collision process and the possibility of related QCD phase transition.
\end{abstract}


\ccode{PACS numbers: 25.75.-q,25.75.Gz, 24.60.Ky, 24.60.Ky, 24.85.+p, 24.10.Lx}

\section{Introduction}
During recent years, numerous attempts have been made to investigate the properties of strongly interacting matter and the existence of the QGP phase produced in heavy-ion collisions under extreme conditions of temperature and baryon-chemical potential~\cite{PT-Jiang,Gazdzicki,Shuryak,Bialas,Appe,Baym,Danilov}. The detection of strongly interacting matter directly in heavy-ion collision experiments is not possible rather the evidence of phase transition is possible by careful examination of the measured extensive quantities like particle multiplicities, mean transverse momentum, particle ratios etc. Lattice QCD calculations~\cite{fedor,ejiri,gavai} predict the existence of QCD critical point and therefore, motivate various theoretical/experimental searches in the field of heavy ion collisions at various places worldwide. The QCD phase transition is characterized by large fluctuations in the observables near the critical point which gets reflected at the final stage in the distribution of measured quantities~\cite{shuryak}. Owing to this fact, correlations and fluctuations of the measured quantities are believed to serve as a thermometer to sense the initial stages of the collision process. Furthermore, the short-range and long-range correlations of the fluctuation observables in separated phase-space intervals are expected to enhance our understanding related to the collectivity effects of produced QCD medium and providing us a signature of the string fusion and percolation phenomenon~\cite{Amelin,Braun1,Braun2}. The informations about the dynamical processes involved in the production of final state charged particles can be extracted by studying fluctuations in particle multiplicity, energy, charge and mean transverse momentum etc.
In this regard, large fluctuations in the charged particle multiplicities can be reliably investigated by the means of suitable (i) extensive and (ii) intensive fluctuation measures. Furthermore, the dependence of fluctuation observables with experimental parameters such as the collision energy ($\sqrt{s_{NN}}$), collision centrality and(or) impact parameter (b), pseudo-rapidity variable ($\eta$), and size of the colliding system (A/B) play a vital role to approach the critical end point and in studying the properties of QCD matter. For example, by varying energy of colliding beams or by varying size of colliding systems, one gets access to vary temperature and/or baryon-chemical potential in a controlled manner. Therefore, studying the behaviour of fluctuation observables with varying  beam energy and target size becomes an obvious choice and has been carried out in present analysis work. During the recent past, a remarkable advancement has been made in fluctuation analysis of high multiplicity data from the heavy-ion collision experiments at CERN-SPS~\cite{e802,wa98,na49,na49-1,na45} and BNL-RHIC~\cite{rhic-1,rhic-2}. Still a lot more theoretical/experimental analysis in this sector is required to be performed. Theoretical model based simulations allow to perform analysis of microscopic properties of system created in nucleus-nucleus (A-B) collisions and provide an opportunity to study multiplicity fluctuations under extreme conditions~\cite{Gorenstein:2008et,Konchakovski:2010fh,Begun:2012wq,Vovchenko:2014ssa,Steinheimer:2016cir,Gorenstein:2011vq,DGhosh,Aduszk,Jeon:2003gk,Asakawa:2015ybt,Konchakovski:2005hq,Konchakovski:2006aq}.\\
Fluctuations observed in extensive quantities such as the  multiplicity ($N_{ch}$) are related to nonthermodynamic fluctuations~\cite{nonthermo} in the initial size of the system and arises mainly due to the distribution of impact parameters, fluctuation in the initial positions of the participating nucleons, and quantum fluctuations arising due to the nucleon-nucleon cross section (fluctuations in the effective size of the nucleons at the initial moment of the collision). All these effects must be taken care of while carrying out the analysis of the experimentally measurable quantities and comparison with the theoretical model predictions should be made. The contribution from the fluctuations in the nucleon-nucleon cross section are comparably small and are not related to transition between QGP and hadronic phase and hence, can be ignored for the analysis purpose. 
The study of multiplicity fluctuations has its limitations because of the finite detector acceptance, detector efficiency and identification of the final state charged particles. In contrast, the average size of the created physical system and their fluctuations on ebe basis are expected to be rather different~\cite{Odyniec,Adare} and strongly affect the observed charged particle fluctuations. Therefore, It is quite necessary to subtract out the system volume fluctuations on ebe basis as these are trivial effects involved in multiplicity fluctuation measurements.
A search of CP may be made by performing a comprehensive two-dimensional scan of phase diagram~\cite{ref57,ref58,ref59} by varying the system size and beam energy~\cite{ref59,ref61}. At CP, the correlation length is expected to increase and makes the fluctuatios as its basic signal~\cite{ref60}. There are indications that the observed fluctuations in the medium size nuclei at SPS energies are related to CP~\cite{ref59,ref60}. These fluctuations are, however, influenced by sereval effects not related to CP or onset of confinement, like conservation laws and volume fluctuations, limited detector acceptance, etc~\cite{ref57,ref59}. Thus, to arrive at some definite conslusion, a systematic analysis of the data should be carried out by using suitable variables so that the contributions from trivial fluctuations may be minimized. The so called intensive variable, $\omega$ is regarded as a suitable one~\cite{ref59,ref61} and is considered in the present study. Recent measurements by the NA61/SHINE collaboration~\cite{Seryakov:2017sss} show  surprisingly interesting features of scaled variances $\omega$ for p-p and nucleus-nucleus A-B collisions.
In this article, we investigate ebe fluctuations in hadron-hadron, hadron-nucleus  collisions at 200 AGeV/c and nucleus-nucleus collisions at  14.6, 60 and 200A GeV/c. Our aim is to study the fluctuation observables by varying size of the pseudorapidity window and the findings are compared with the predictions of the Lund Monte Carlo based FRITIOF model. \\
The paper is organized in the following manner: Section II provides a short description of the experimental data used for analysis. In section III, the description of FRITIOF model is provided, section IV presents the formalism adopted for the analysis carried out. The analysis results are discussed in Section V, whereas, the summary of results followed by discussion is presented in section VI.

\section{Details of the Experimental Data}
The present analysis work is carried out using sample of interactions as obtained from the exposure of the NIKFI BR-2 nuclear emulsion stacks to $^{16}$O beam with momenta 14.6A, 60A and 200A GeV/c and to proton beam having momenta 200A GeV/c. These experiments were performed by EMU-01 collaboration~\cite{Bebecki,Gurtu,EMU-1,EMU-2,EMU-4}. All the relevant details about the data used in the present study, like scanning procedure, selection criteria for events, classification of tracks, sorting of the H, CNO and AgBr groups of interactions, method of measuring the space angles of relativistic charged particles etc. may be found elsewhere~\cite{busra,ahmad1,ahmad2,ahmad3}. It should be mentioned that conventional emulsion technique have two main advantages over the other detectors: i) its 4$\pi$ solid angle coverage and ii) data are free from biases due to full phase space acceptance.\\
The details of the event samples used in the present study are listed in Table 1.
\begin{table}[ht]
\caption{Number of interaction events selected for the present
analysis.} 
\centering 
\begin{tabular}{|c| c| c| c|} 
\hline\hline 
Beam Momentum (A GeV/c) & Type of interactions & No. of events \\ [0.5ex] 
\hline 
 200 & p-H & 242 \\ 
 200 & p-CNO & 655 \\
 200 & p-AgBr & 855 \\
 14.6 &O-AgBr & 379 \\
 60 & O-AgBr & 422 \\
 200 & O-AgBr & 223 \\ [1ex] 
\hline 
\end{tabular}
\label{table:nonlin} 
\end{table}

\section{The FRITIOF Model}
FRITIOF is a Monte Carlo program that implements the Lund string dynamics model for hadron-hadron, hadron-nucleus and nucleus-nucleus interactions at relativistic energies~\cite{FRI-1,FRI-2,FRI-4}. The model considers the collision between two nuclei as nucleon-nucleon collisions, nuclei being the composition of various constituent entities. The final multiparticle state in these collisions is assumed to be  the result of the fragmentation of the longitudinally excited strings created between the interacting nucleons. Fragmentation process of these longitudinally excited strings is responsible for the production of charged particles in the final stage of the collision.
Several other models based event generators like AMPT, EPOS, UrQMD etc are being used for the comparison of experimental measurements with theoretical predictions but the modified FRITIOF code~\cite{FRI-1,FRI-2} is preferred in the present analysis work as it is suitable to mimic the measurements performed by nuclear emulsion experiments at relativistic energies. The proportional abundance of different categories of target nuclei present in the emulsion material is properly implemented and the modified code takes into account the percentage of interactions occurring with different targets of emulsion like H, C, N, O, Ag, Br, etc~\cite{DGhosh}. In order to compare the experimental findings with the model, events corresponding to each considered real data set were simulated using FRITIOF model code. The number of events in each sample is kept nearly the same as that in the corresponding experimental events.
\section{Formalism}
The critical point is characterized by the increase in the spatial correlation length $\xi$. The large spatial correlation length $\xi$ causes larger number of particles being correlated which gets reflected in the form of a larger magnitude of the fluctuation measures~\cite{Seryakov:2017sss}. These fluctuation measures are related to the thermodynamic properties of the system, like, entropy, specific heat, chemical potential, etc and would help understand the nature of phase transition and the critical fluctuations at the QCD phase boundary. In this regard, the investigation is carried out by measuring the pseudorapidity dependence of fluctuation observables. Comparative analysis of elementary p-p interactions is utilized as the baseline for making the investigations of nucleus-nucleus collisions more compact and complete as it deciphers the high-energy A-B collisions at similar energies. In order to eliminate the trivial volume fluctuations or unfavorable detector effects like inefficiency of the detector, we calculate intensive and strongly intensive variables for studying the critical behavior.\\
  The mean multiplicity and the variance of multiplicity distribution are defined as:
     \begin{equation}
       {\left<N_{ch}\right>}= \sum{N_{ch}P(N_{ch})},
   \label{eqn:average}
   \end{equation}
 \begin{equation}
   Var(N_{ch}) = \sum(N_{ch} -\left<N_{ch}\right>)P(n)={\left<N_{ch}^2\right> - \left<N_{ch}\right>^2},
   \label{eqn:variance}
 \end{equation}
 where, $N_{ch}$ denotes the charged particle multiplicity per event while $<..>$ gives the average value of the quantity. \\
 Within the grand canonical ensemble, mean and variance of a multiplicity distribution are extensive quantities as they are proportionl to system volume (V). Whereas, the scaled variance $\omega$,  the ratio of two extensive quantities (the variance and mean) is an intensive variable and is expressed as~\cite{Ghosh}:\\
\begin{equation}
  \omega = \frac{Var(N_{ch})}{\left<N_{ch}\right>}=\frac{\left<N_{ch}^2\right> - \left<N_{ch}\right>^2}{\left<N_{ch}\right>}\,
  \label{eqn:omega}
\end{equation}
In the context of heavy-ion collisions, the scaled variance~\cite{Ghosh,alt} is utilized  to reveal the nature of the correlation among the produced particles as it can be interpreted as an aspect of a two-particle correlation function. Inaccuracy in determining the centrality in symmetric/asymmetric A-B collisions give rise to event-by-event volume fluctuations. From one collision event to another, the size of the system can not be kept fixed due to the variation of impact parameter. Moreover, for a given impact parameter, the number of participating nucleons would vary on ebe basis which, in turn, would influence the scaled variance. Although the scaled variance does not depend on system volume but is sensitive to the volume fluctuations. It is, therefore preferred to to measure the scaled variance in limited collision centrality bins, e.g., 0.5\% most central or 50-60\% central collisions~\cite{Seryakov:2017sss} so that it's effects observed may be mostly of dynamical origin.\\
 In order to eliminate the system volume effect of imperfectly known distributions, the strongly intensive variables $\Sigma_{A-B}$\cite{Gorenstein:2011vq} is considered. It is combinations of cumulants in which the spatial volume fluctuations are canceled out. 
Taking the extensive event quantities $A$ as the charged particle multiplicity $N_{F}$ in the Forward pseudorapidity window and $B$ as the multiplicity $N_{B}$ in the Backward pseudorapidity window, $\Sigma_{FB}$ is defined as:
\begin{eqnarray}
  \Sigma_{FB}=\frac{1}{C_\Sigma}\left[\langle N_{B}\rangle\omega[N_{F}]+\langle N_{F}\rangle\omega[N_{B}]-2\cdot (\langle N_{F}\cdot N_{B}\rangle-\langle N_{F}\rangle\langle N_{B}\rangle)\right]
  \label{eqn:Sigma}
\end{eqnarray}
where,
\begin{eqnarray}
 \omega[N_F]=\frac{\langle{N^{2}{_F}} \rangle - {\langle{N{_F}} \rangle}^2}{\langle{N{_F}} \rangle},
 \omega[N_B]=\frac{\langle{N^{2}{_B}} \rangle - {\langle{N{_B}} \rangle}^2}{\langle{N{_B}} \rangle},
\end{eqnarray}
and
\begin{eqnarray}
 C_\Sigma ={\langle{N{_B}} \rangle + {\langle{N{_F}} \rangle}} 
\end{eqnarray}\\

In Eq.~\ref{eqn:Sigma}, the quantities, $\langle N_{F} N_{B}\rangle$ and $\langle N_{F}\rangle \langle N_{B}\rangle$, are the forward-backward multiplicity correlation terms.
$\Sigma_{FB}$ is a two dimensional second order moment and is directly related to thermodynamics susceptiblities of a physical system~\cite{Gorenstein:2011vq,evan}. In the context of independent source models, strongly intensive quantity $\Sigma_{FB}$, should carry information related only to the characteristics of single average source producing particles and the analysis can reveal information on the early dynamics of the nuclear collisions at relativistic energies~\cite{iwona}.
\section{Results}
\subsection{Multiplicity and Pseudo-rapidity distributions}
\begin{figure}
\centering
\resizebox{1.0\textwidth}{!}{
\includegraphics{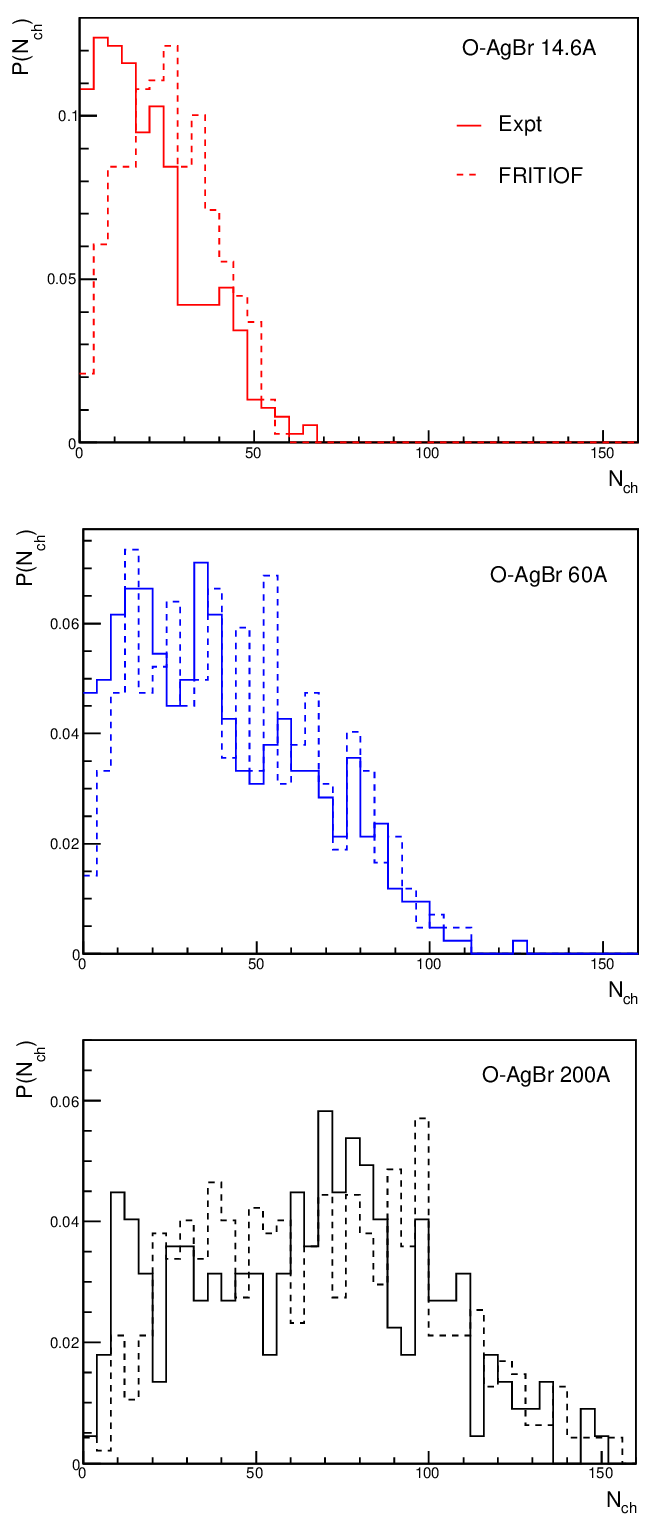}
\includegraphics{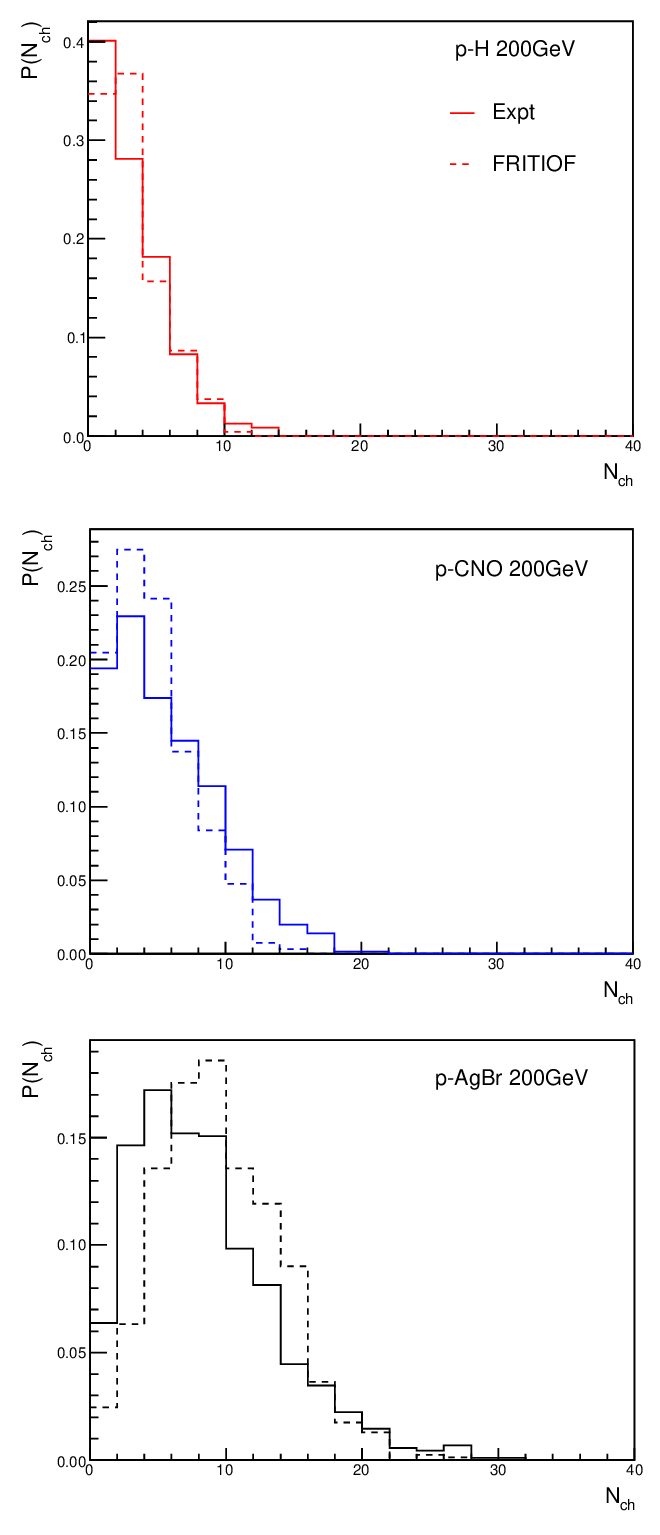}
 }
\caption{The experimentally measured charged particle multiplicity distributions for O-AgBr collisions at 14.6A, 60A, and 200A GeV/c (left panel) and for proton induced collisions with H, CNO and AgBr target groups at 200A GeV/c (right panel) presented by solid lines. The corresponding simulation results obtained from the Lund Monte Carlo based FRITIOF model is presented by broken lines.}
\label{fig:Pch}
\end{figure}

\begin{figure}
\centering
\resizebox{1.0\textwidth}{!}{
\includegraphics{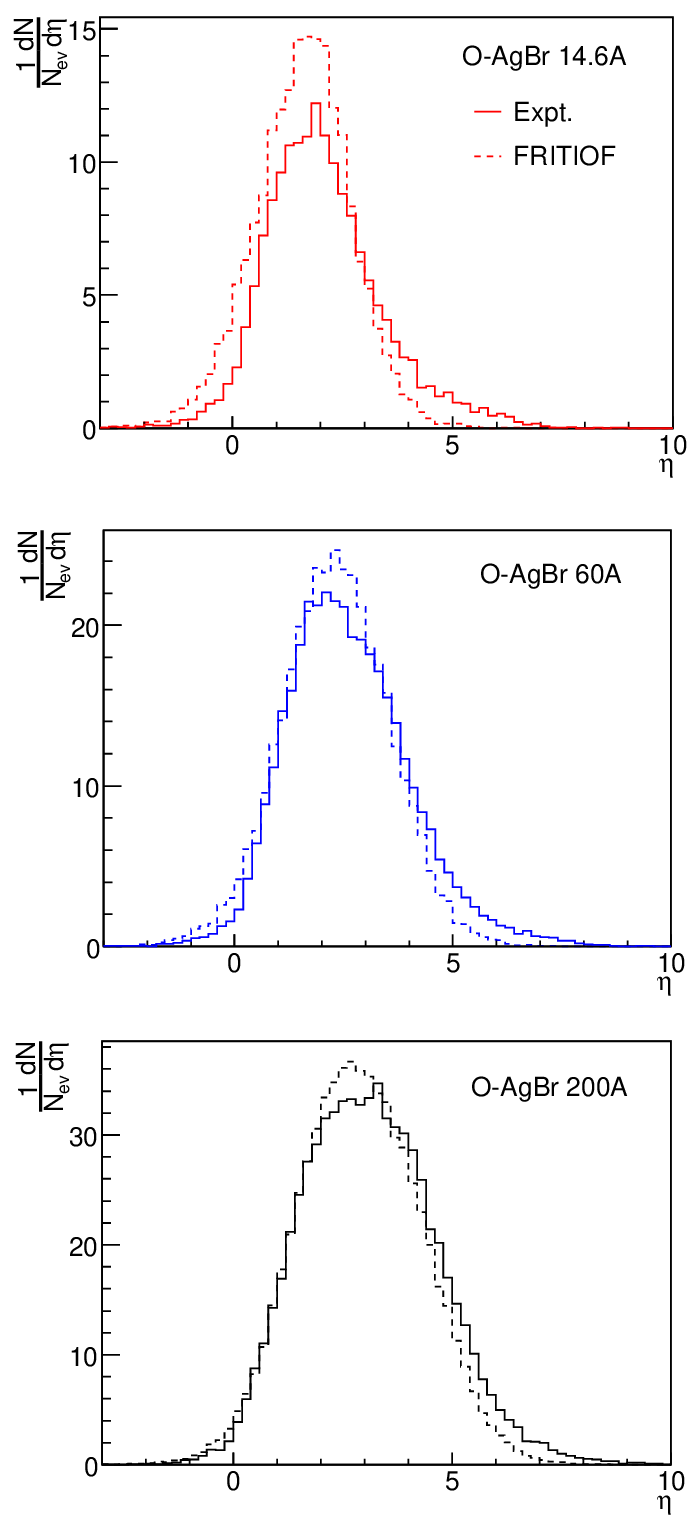}
\includegraphics{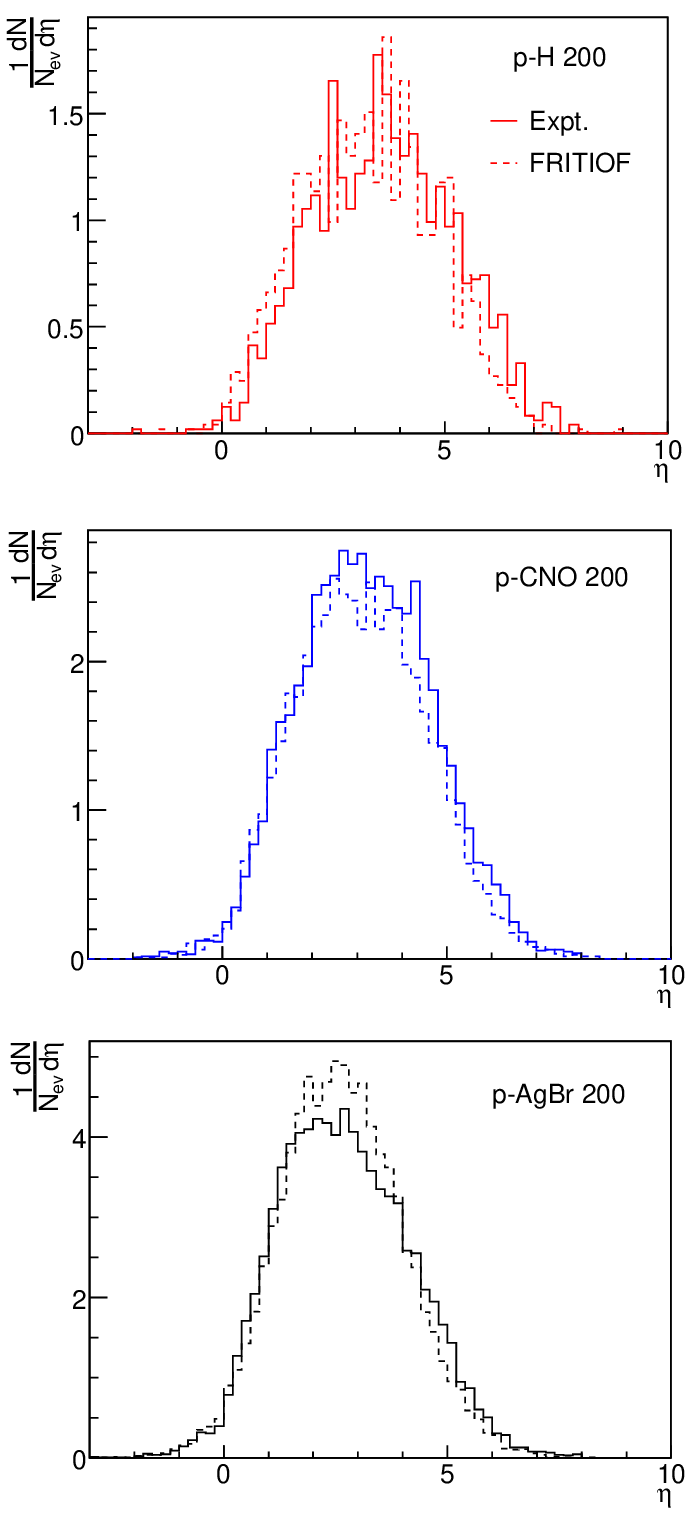}
 }
\caption{Charged particle pseudorapidity distributions for O-AgBr collisions at 14.6A, 60A, and 200A GeV/c (left panel) and for proton induced collisions with H, CNO and AgBr targets 200A GeV/c (right panel). The broken lines correspond to FRITIOF simulated results.}
\label{fig:pseudo}
\end{figure}

Multiplicity  distributions for O-AgBr collisions at 14.6A, 60A, and 200A GeV/c and for proton induced collisions with H, CNO and AgBr targets at 200A GeV/c in  are displayed in Figure~\ref{fig:Pch}. Multiplicity distributions obtained from the FRITIOF simulated events are also presented in the same figure.
Figure~\ref{fig:pseudo} shows the pseudorapidity distributions for O-AgBr collisions at three beam momenta and for proton induced collisions with different target nuclei groups at 200A GeV/c. $\eta$-distributions corresponding to FRITIOF data are also plotted in these figures.

\subsection{Mean and Variance of particle multiplicities}
Two-particle correlations in p-p collisions at the AGS, SPS, RHIC and LHC energies has been widely used as a tool to estimate the influence on multiplicity correlations of different particle sources in heavy-ion collisions and to understand the mechanism of multiparticle production. In this section, charged particle multiplicity, $N_{ch}$ has been studied as a function of $\eta$ window width in the $\eta$ interval $-3.0 \le \eta_c \le 3.0$.
The mean value of fluctuating extensive quantities (for example, $N_{ch}$) from one event to another and its variance $\sigma$, both are proportional to the system volume in the asymptotic limit of large system volumes and are used to characterize the mutliparticle production mechanism. 
\begin{figure}
\centering
\includegraphics[scale =0.65]{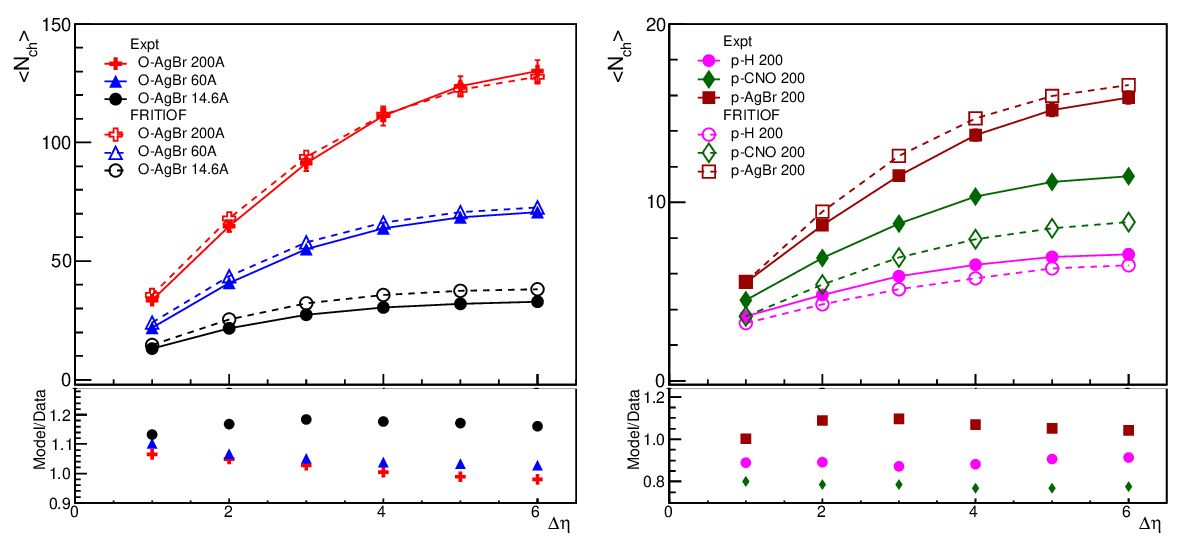}
\caption{Variations of $\left\langle N_{ch}\right\rangle$ as a function of $\Delta \eta$ for O-AgBr collisions at 14.6A, 60A and 200A GeV/c~\cite{EMU-1,EMU-2,EMU-4}(left panel) and for proton beam induced collisions with H, CNO and AgBr at 200A GeV/c (right panel) alongwith the FRITIOF simulations. The error bars shown in the plot are statistical uncertainties only. The ratio plots, model/data, are shown in the bottom panels.}
\label{fig:nch}
\includegraphics[scale =0.65]{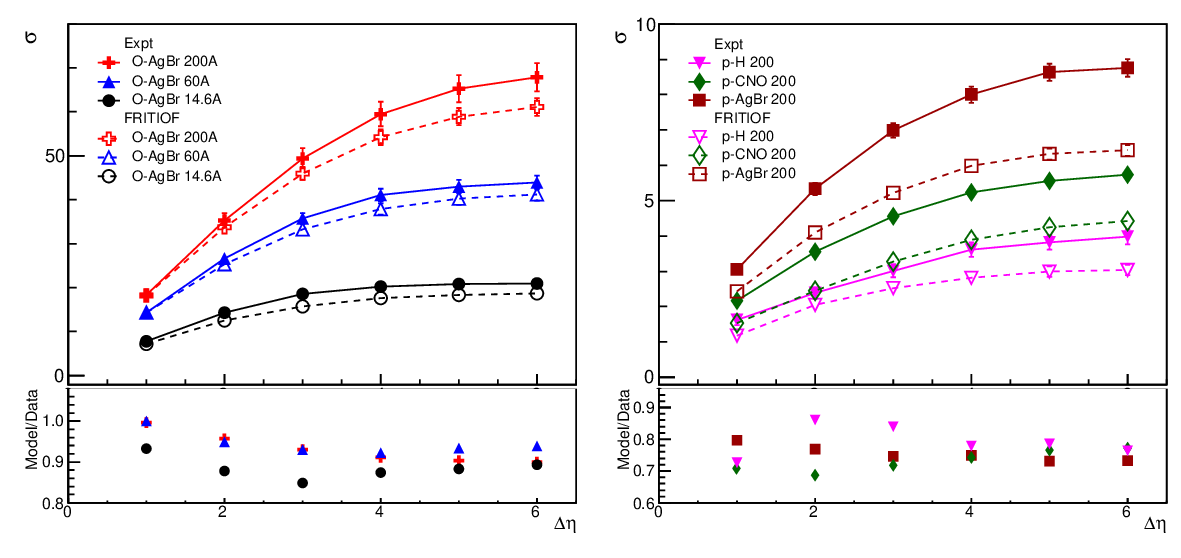}
\caption{Variations of $\sigma$ on $\Delta \eta$ for O-AgBr collisions (left panel) and for proton beam induced collisions (right panel)~\cite{EMU-1,EMU-2,EMU-4} alongwith the predictions of FRITIOF model. The error bars shown in the plot are statistical uncertainties only. The ratio plots, model/data, are shown in the bottom panels.}
\label{fig:sigma}
\end{figure}

In Figures ~\ref{fig:nch} and ~\ref{fig:sigma}, the dependence of mean multiplicity, $\langle{N_{ch}}\rangle$ and variance of the multiplicity distribution, $\sigma$ on the widths of the $\eta$-windows are shown for real and the corresponding FRITIOF simulated data sets (p-H, p-CNO and p-AgBr at 200A GeV/c and O-AgBr interactions at incident beam momenta 14.6A, 60A and 200A GeV/c. For this, a pseudorapidity window of width $\Delta \eta =1.0$ is chosen and placed in such a way that its center coincides with the center of symmetry of $\eta$ distribution. Thus, all the charged particles having their $\eta$ values in the interval $\eta_c \pm \Delta\eta/2$ are counted to estimate the mean charged particle multiplicity, $N_{ch}$, $\sigma$, $\omega$, etc. in this window. The window width is then increased in steps of 1.0 until a region $\eta_{c} \pm 3.0$ is covered. It is observed in Figure~\ref{fig:nch} that for all the data sets, the mean multiplicity grows with increasing width of pseudo-rapidity window. The growing trend of $N_{ch}$ values with $\Delta \eta$ for the experimental data and those from the simulation using FRITIOF code are qualitatively in good agreement, whereas numerical values are slightly under predicted by the model, as is evident from the figure. The observed dependence of $\sigma$ on $\Delta \eta$, shown in Figure~\ref{fig:sigma} may be explained in the context of clusters: with increasing $\Delta \eta$, the probability that more than one particle arising from a single cluster may fall in either of the two regions increases.
In the absence of correlated emission of particles, the variance is predicted~\cite{DGhosh} to be $\sim$ 1. Significantly higher values of $\sigma$ than unity, observed, with increasing $\eta$-intervals, are noticeable in the figure. Such a deviation from unity in the $\sigma$ values has also been reported at RHIC energies~\cite{Haussler,Wozniak-3}. Significantly larger $\sigma$ values for O-AgBr interactions as compared to those for p-AgBr interactions can be interpreted in terms of higher number of nucleonic collisions due to lager projectile mass. The bottom panels of Figures~\ref{fig:nch} and~\ref{fig:sigma} show the ratio plots of model predicted values to the experimental data as a function of $\Delta \eta$. It is evident from the ratio plot that for O-AgBr collisions, the numerical values of $N_{ch}$ are slightly overestimated by the model. For proton induced collisions, on the other hand, the model predicted $\left\langle N_{ch}\right\rangle$ values are somewhat smaller as compared to those obtained from the real data except for p-AgBr collisions where model predicted values are larger than those from the experimental data. From Figure~\ref{fig:sigma}, it may be noted that the model predicted $\sigma$ values are smaller than those from the real data for all data sets. \\

\subsection{Scaled variance $\omega$, $F_2$ and $f_2$ moments}
\label{sec:3}
 The scaled variance $\omega$, although cancels out the system volume effects~\cite{Gorenstein:2011vq}, yet it is sensitive to the system volume fluctuations. If there are no fluctuations present, the normalization condition results $\omega=0$ and indicates that there is no correlation amongst the produced charged particles. If charged particles were produced independently, the multiplicity distributions would be Poissonian and the variance will be equal to mean of the distribution, resulting $\omega= 1$. If there is correlated emission of the charged particles, it would be inferred in terms of a deviation of the multiplicity distribution from the Poissonian resulting the value of $\omega$ different from unity. It should be mentioned that in the vicinity of  critical end point, the scaled variance  $\omega$ of the multiplicity distributions are expected to diverge~\cite{Adare,Vovchenko} and therefore are regarded as, quite useful in studying the anomalous behavior of strongly interacting matter.

\begin{figure}
  \includegraphics[width=13cm, height=6.5cm]{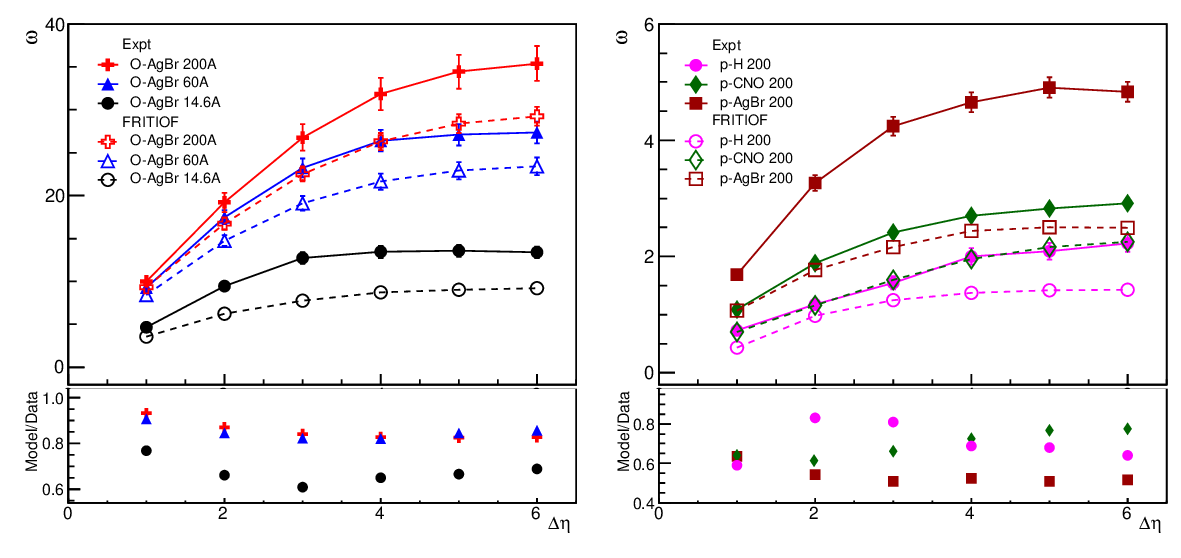}
  \caption{Variations of the scaled variance, $\omega$ with $\Delta \eta$ for O-AgBr collisions (left panel) and for the proton induced collisions (right panel)~\cite{EMU-1,EMU-2,EMU-4}. The error bars shown in the plot are statistical uncertainties only. The ratio plots, model/data, are shown in the bottom panels.}
  \label{fig:fig5}
\end{figure}

Variations of $\omega$ with $\Delta \eta$ for the real and FRITIOF data are presented in Figure~\ref{fig:fig5}. It is reflected from the figure that for proton-nucleus interactions at 200A GeV/c, $\omega$ increases monotonically upto a $\Delta \eta$ $\sim$ 2 and thereafter tend to acquire a saturation. The values of $\omega >> 1$ clearly indicates the presence of correlated emission of charged particles. Maximum values of scaled variance indicate maximum correlation and the corresponding phase space region corresponding to the region of maximum correlation. The results from the real data sets, thus, indicate the presence of correlations among the produced charged particles, FRITIOF simulated results too indicate the presence of correlations but with somewhat smaller magnitudes.

\begin{figure}
\includegraphics[width=13cm, height=6.5cm]{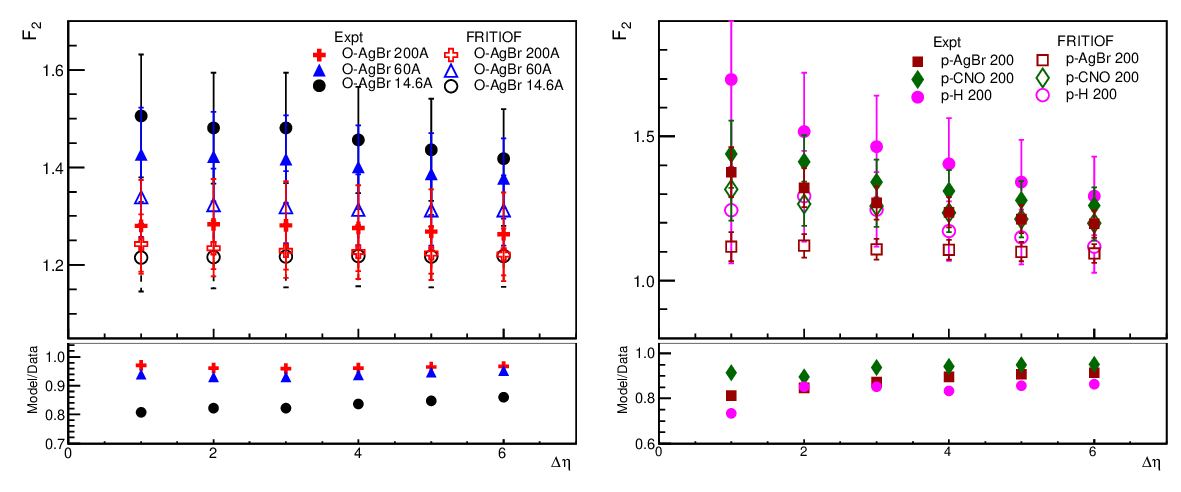}
\caption{Variations of $F_{2}$ on $\Delta \eta$ for O-AgBr collisions (left panel) and for proton induced collisions (right panel)~\cite{EMU-1,EMU-2,EMU-4} alongwith the FRITIOF simulations. The error bars shown in the plot are statistical uncertainties only. The ratio plots, model/data, are shown in the bottom panels.}
\label{fig:F2}
\end{figure}
   
\begin{figure}
  \includegraphics[width=13cm, height=6.5cm]{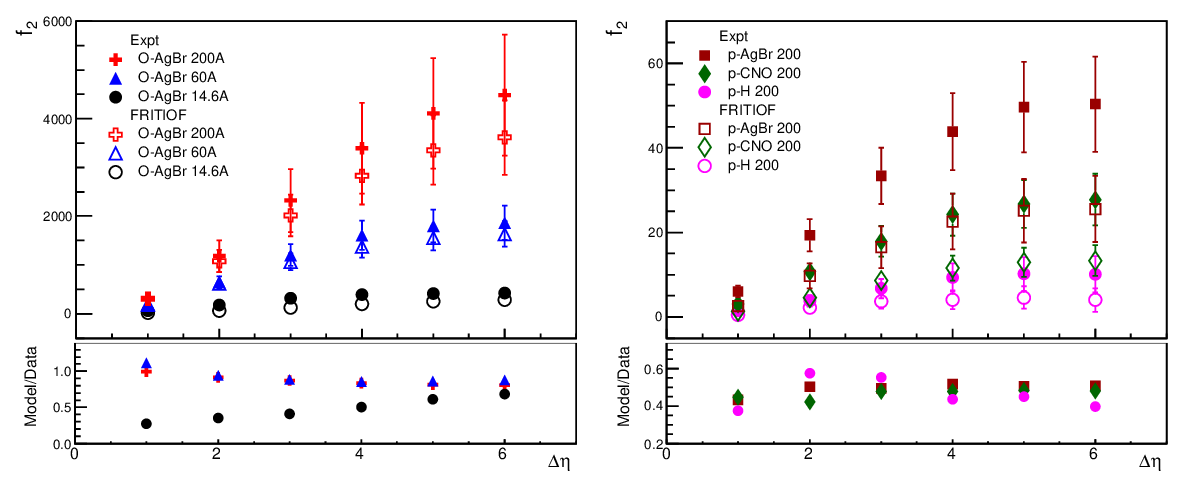}
  \caption{Variations of $f_2$ as a function of $\Delta \eta$ for various data sets. The error bars shown in the plot are statistical uncertainties only. The ratio plots, model/data, are shown in the bottom panels.}
  \label{fig:f2}
  \vspace{-0.1cm}
\end{figure}

The quantity $\omega$ defined by the equation~\ref{eqn:omega}, can be plagued by efficiency. The scaled factorial moments (SFMs) can be a better choice to cancle the efficiency effects. The method of SFMs~\cite{ref48,ref49} have been widely used~\cite{busra,ahmad3,ref50,ref51,ref52} to search for the non-linear phenomenon in hadronic and heavy-ion collisions. SFMs has also been used to study the various processes at SPS and RHIC energies~\cite{busra,ref53} with the aim to scan the phase diagram in a systematic search for QCD critical point. The scaled factorial moment of order, q=2 is given by
\begin{eqnarray}
F_2 = \frac{\left\langle N_{ch}(N_{ch}-1)\right\rangle}{\left\langle N_{ch}\right\rangle^{2}}
\end{eqnarray}
Variations of $F_{2}$ with $\Delta \eta$ for the experimental and FRITIOF data samples are plotted in Figure~\ref{fig:F2}. The ratio plots, FRITIOF/data are also displayed in the bottom panels of the figure. It may be noted from the figure that $F_2$ gradually decreases with the increasing width of the $\eta$ windows. Furthermore, for a given $\Delta \eta$, $F_2$ decreases with increasing beam energy and(or) target size. FRITIOF model also predicts almost similar trend as those exhibited by the data but with smaller magnitudes. Smaller values of $F_{q}$ moments against $\Delta\eta$ for the FRITIOF model as compared to experimental values have also been observed by EMU01 collaboration~\cite{ref53a}. Such a difference in the experimental and FRITIOF values may arise due to the difference in the multiplicity distribution for the real and FRITIOF data with a given $\Delta\eta$. Any factorial moment of a given order contains information about all the lower orders of multiparticle correlation and is dominated by two particle correlation~\cite{ref50}. Therefore, in order to focus on the correlation of a given order, these lower order correlations are to be removed. To account for this, factorial cumulants were introduced~\cite{ref50,ref54} which are estimated by subtracting the influence of all lower order moments from a given factorial moments~\cite{ref50,ref55}. The factorial cumulant of order, q=2 given by
\begin{eqnarray}
f_2 = \left\langle N_{ch}(N_{ch}-1)\right\rangle -\left\langle N_{ch}\right\rangle^{2}
\end{eqnarray}
are obtained by subtracting the first order contribution, $f_1 = \left\langle N_{ch}\right\rangle^{2}$ from the second-order factorial moment~\cite{ref50,ref54,ref56}. For a poisson distribution, which characterizes the independent emission of particles or for a coherent emission $f_2 = 0$, whereas for chaotic sources $f_2 > 0$. Variations of $f_2$ with $\Delta\eta$ for various data sets considered in the present study (real and simulated) are shown in Figure~\ref{fig:f2}. It is observed that the values of $f_2$ increase with widening of $\eta$ window width upto $\Delta\eta \sim 5$ and thereafter tend to acquire a saturation. It may also be noted that for a given $\Delta\eta$, $f_2$ increases with beam energy and target size. Model predicted values of $f_2$ are also found to follow nearly similar trend of variation except that the model predicts smaller values as compared to the real data. These findings are, thus, in close agreement with those obtained in minimum bias and central S-S collisions at 200 GeV~\cite{ref56}.  

 \begin{figure}  
 \centering 
\includegraphics[width=13cm, height=6.5cm]{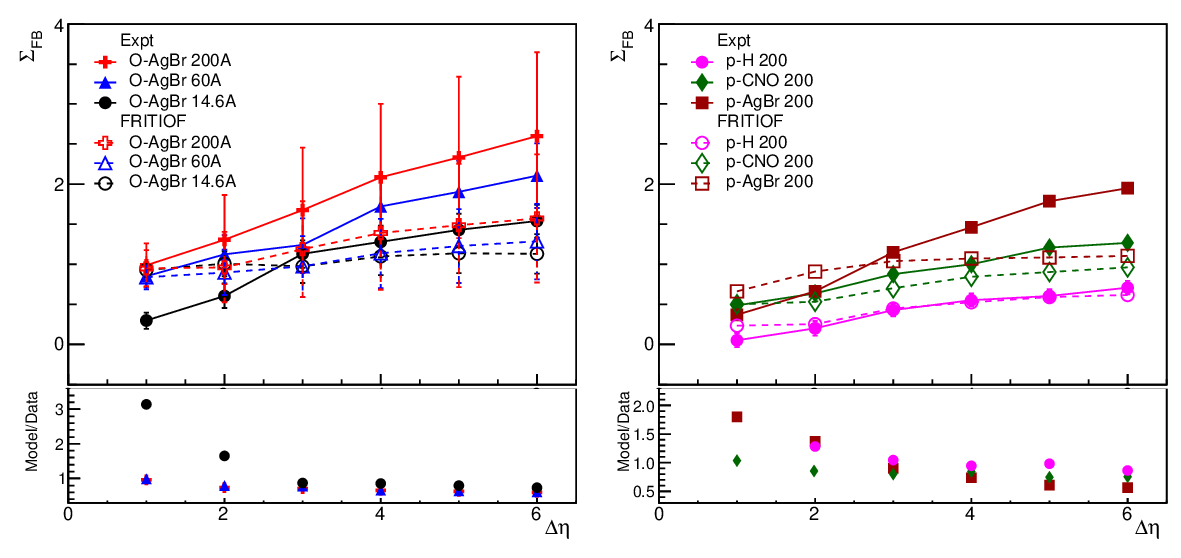}
\caption{Variations of $\Sigma_{FB}$ on $\Delta \eta$ for experimental and FRITIOF data samples. The error bars shown in the plot are statistical uncertainties only. The ratio plots, model/data, are shown in the bottom panels.}
\label{fig:Sigma}
\end{figure}

 \begin{figure}  
 \centering 
\includegraphics[width=8cm, height=14cm]{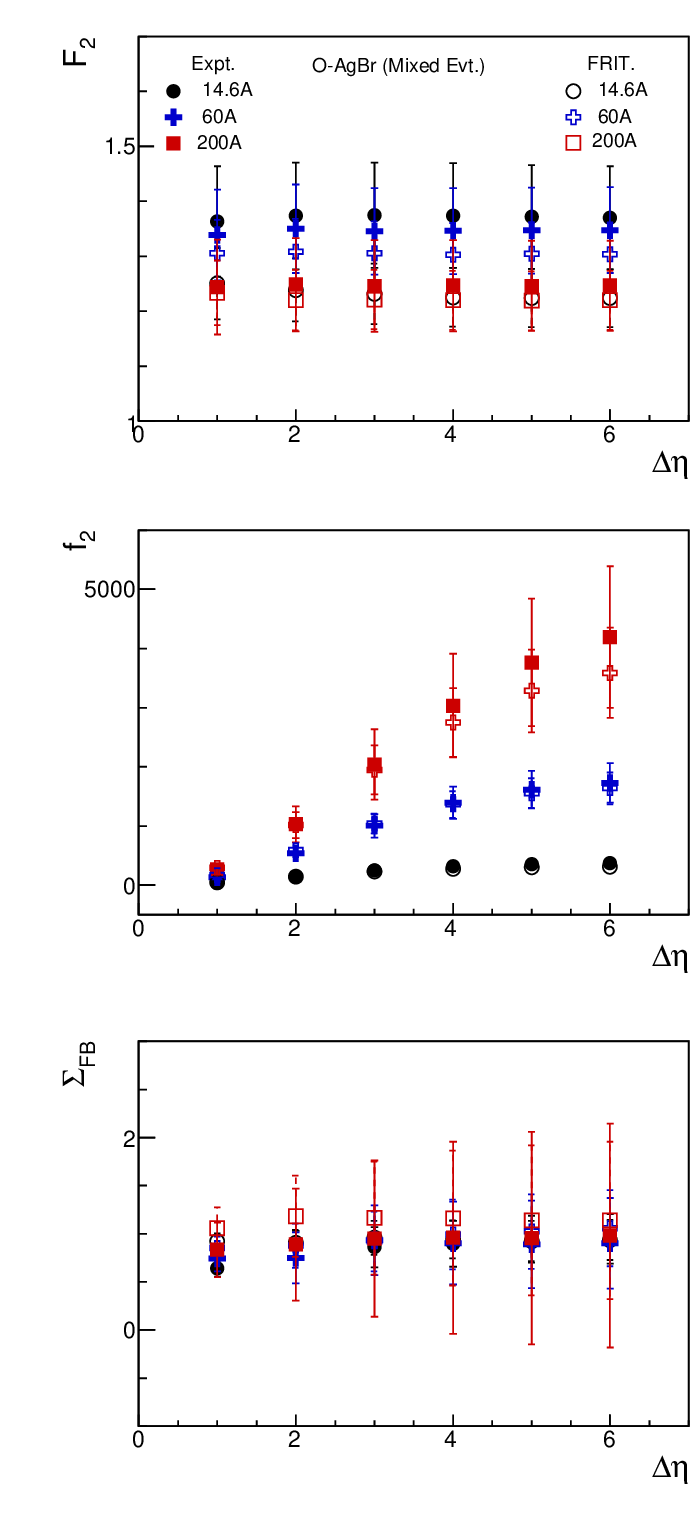}
\caption{Variations of $F_2$, $f_2$ and $\Sigma_{FB}$ with $\Delta\eta$ for the mixed events corresponsing to the real and FRITIOF data at three beam energies.}
\label{fig:Sigma_mxd}
\end{figure}

\subsection{Strongly intensive variable $\Sigma_{FB}$}
\begin{figure}
\centering
\resizebox{0.95\textwidth}{!}
{
\includegraphics{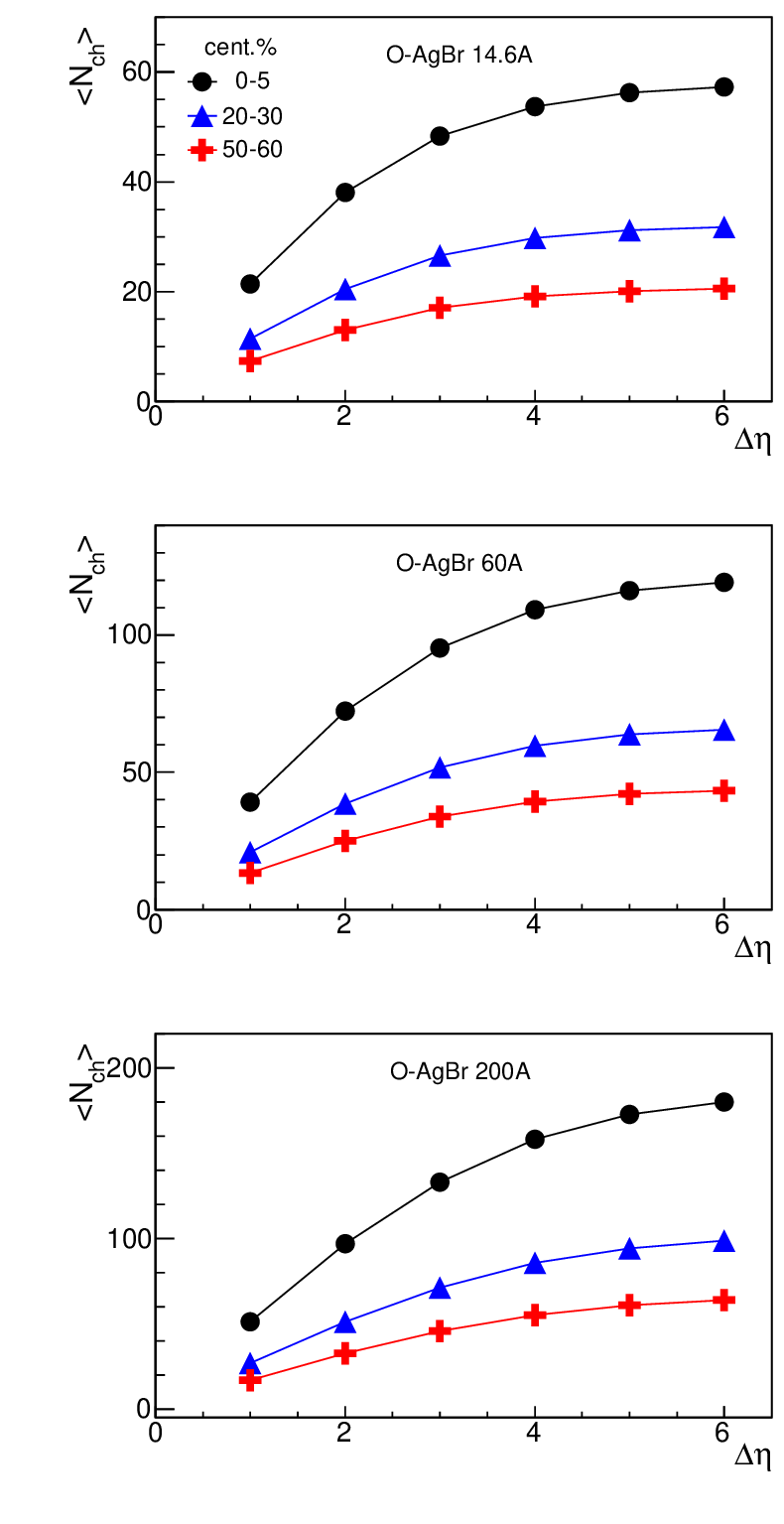}
\includegraphics{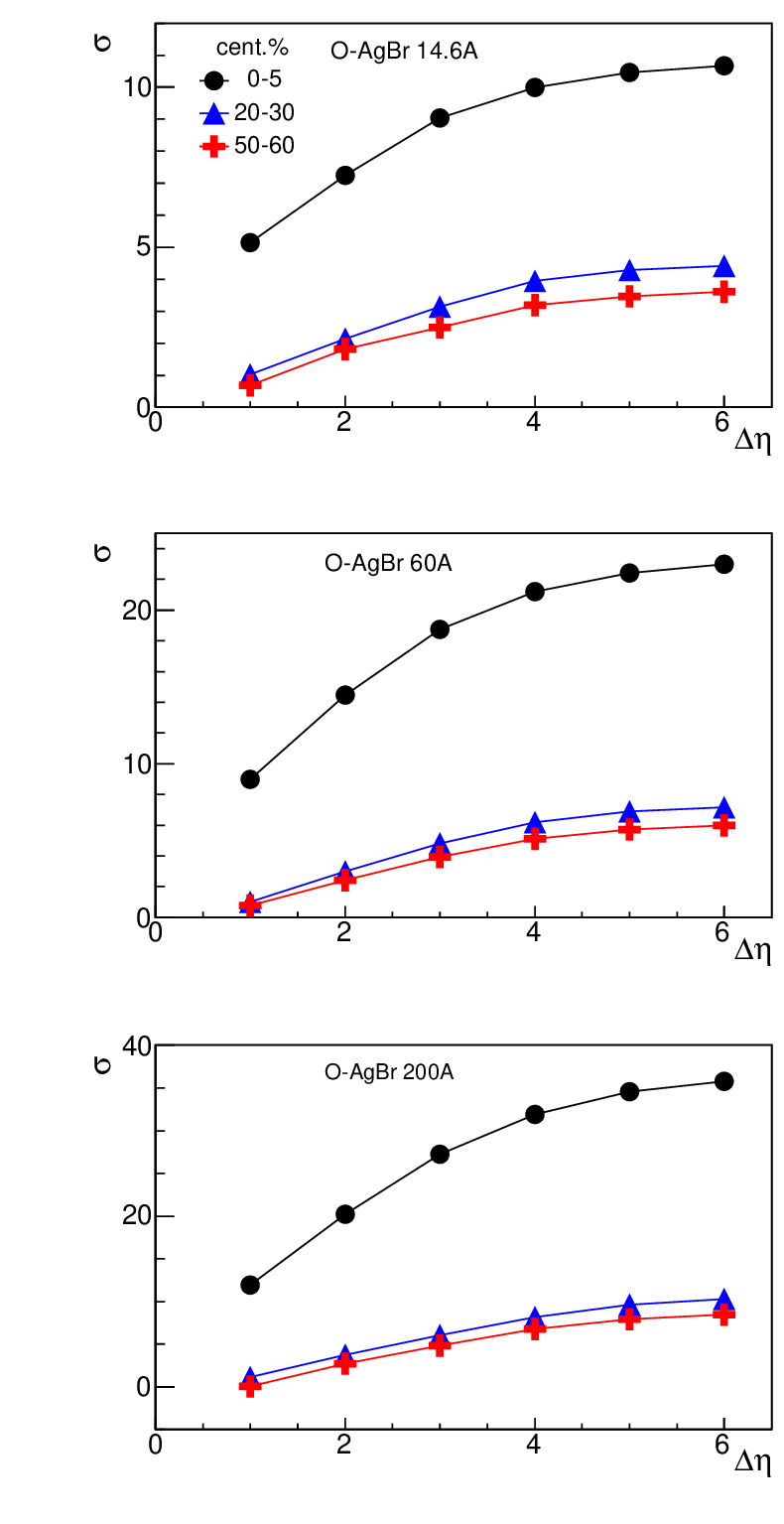}
 }
\caption{Variations of $\left\langle N_{\rm ch}\right\rangle$(left panel) and $\sigma$(right panel) on $\Delta \eta$ for the centrality classes 0-5\%, 20-30\% and 50-60\% for FRITIOF simulated events corresponding to O-AgBr collisions at 14.6A, 60A and 200A GeV/c. The error bars shown in the plot are statistical uncertainties only.}
\label{fig:nch_cent}
\end{figure}

\begin{figure}
\centering
\includegraphics[width=8cm, height=14cm]{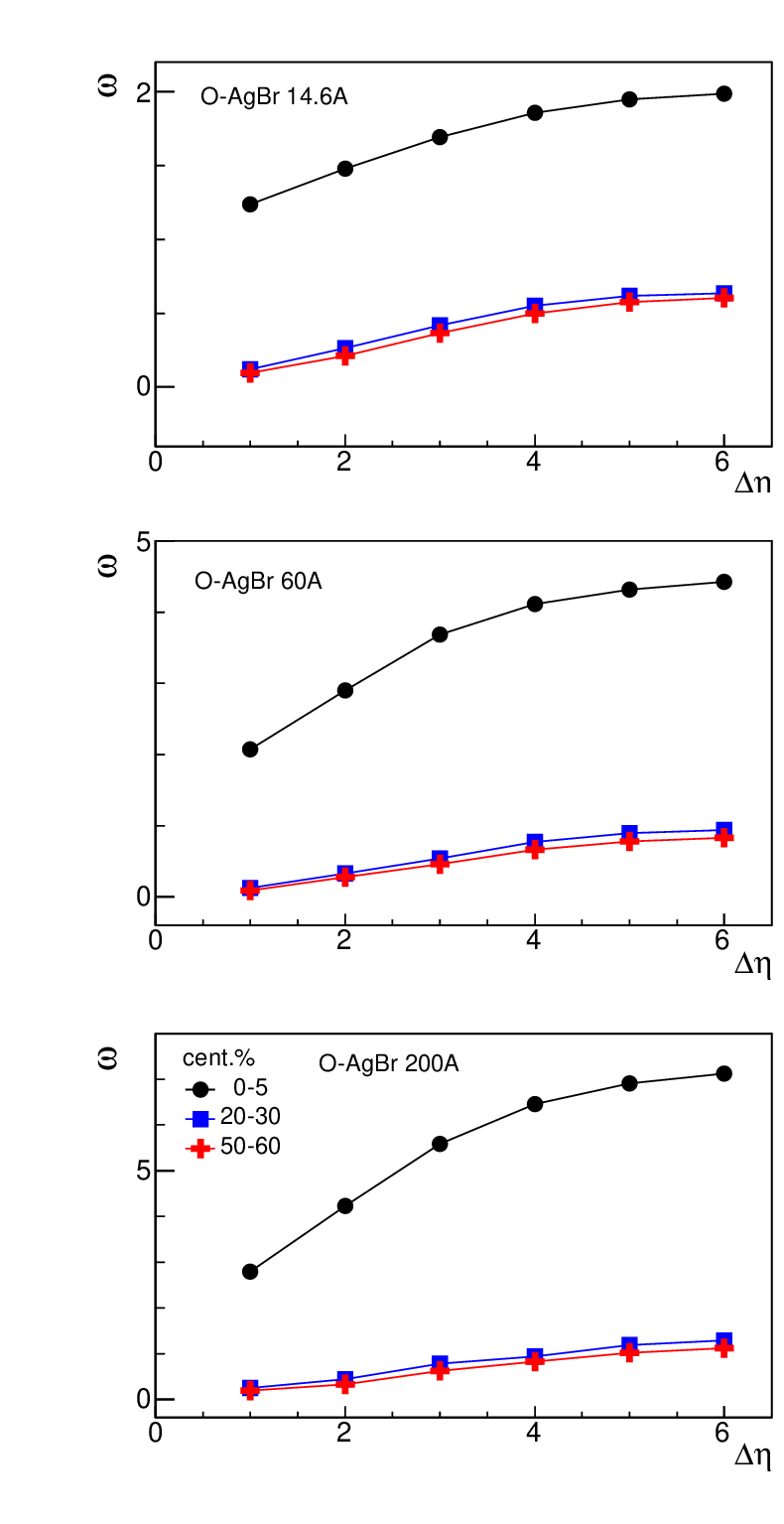}
\caption{Variations of $\omega$ on $\Delta \eta$ for the three centrality classes for FRITIOF simulated O-AgBr collisions at different momenta. The error bars shown in the plot are statistical uncertainties only.}
\label{fig:omega_cent}
\end{figure}

\begin{figure}
\centering
\includegraphics[width=8cm, height=14cm]{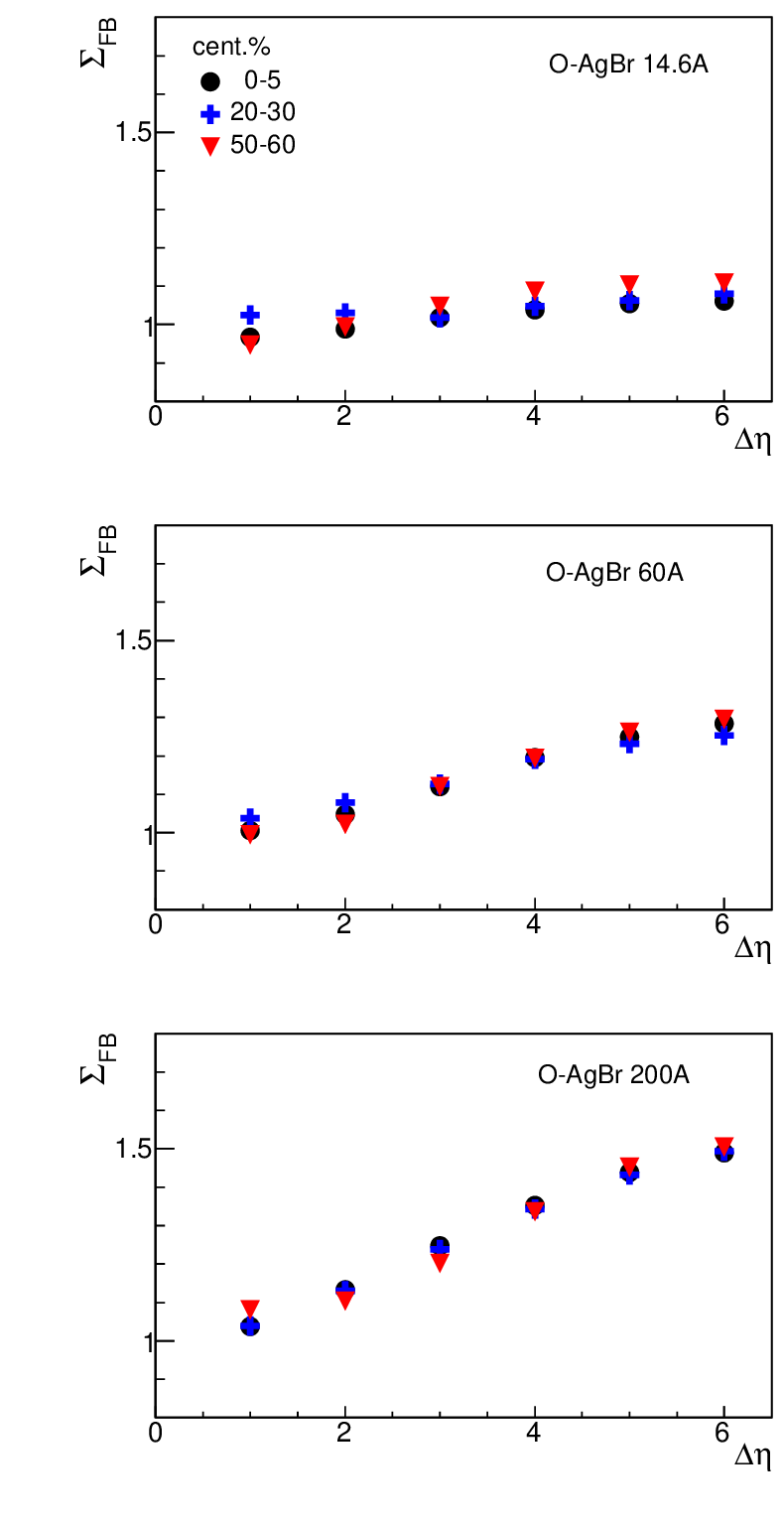} 
\caption{Variations of $\Sigma_{FB}$ on $\Delta \eta$ for various centrality classes for FRITIOF simulated events corresponding to 14.6A, 60A and 200A GeV/c O-AgBr collisions. The error bars shown in the plot are statistical uncertainties only.}
\label{fig:sigma_cent}
\end{figure}

\begin{figure}
\centering
\includegraphics[width=8cm, height=14cm]{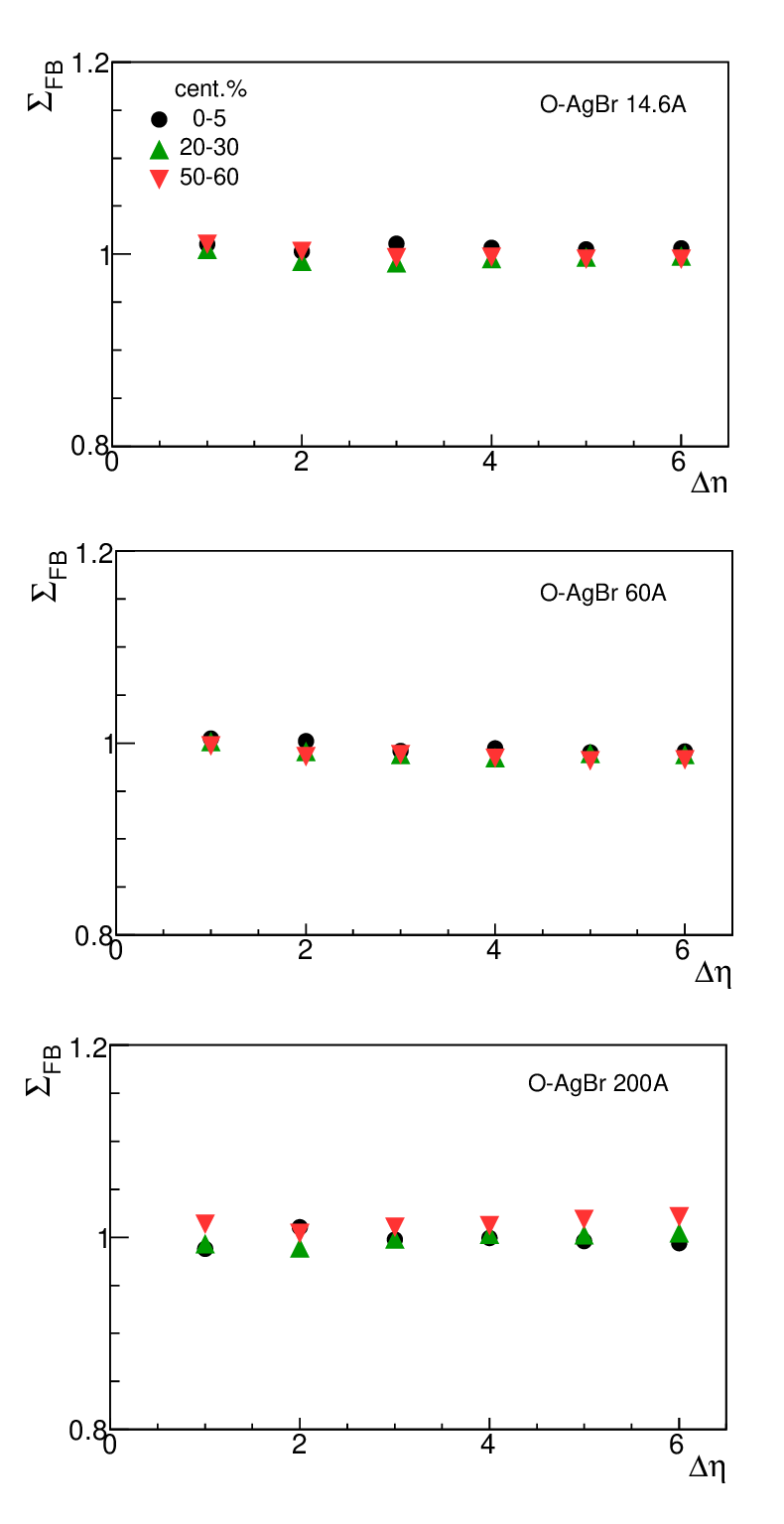} 
\caption{Variations of $\Sigma_{FB}$ on $\Delta \eta$ for the mixed events for the three centrality selections at the three beam energies.}
\label{fig:sigma_mxd_cent}
\end{figure}

Variation of $\Sigma_{FB}$ with the increasing  width of the pseudo-rapidity interval for the real and FRITIOF simulated data sets are displayed in Figure~\ref{fig:Sigma}. It may be noted from figure that $\Sigma_{FB}$ values show the slow rising trend with increasing width of the pseudo-rapidity interval for FRITIOF events indicating a weak short range correlation for p-H and p-A collisions (right panel). The real data sets, however, show a rather stronger rising trend. It is interesting to note that $\Sigma_{FB}$ acquires a value $> 1$ for p-AgBr collisions for $\Delta\eta \gtrsim 2$. For O-AgBr collsions, $\Sigma_{FB}$ is observed to be $> 1$ for $\Delta\eta \gtrsim 2$ and goes on increasing with widening of $\eta$-windows and/or increasing beam momenta. It should be mentioned here that the basic properties of $\Sigma_{FB}$ are~\cite{ref1}:\\
$\Sigma_{FB}$ = 0 in the absence of fluctuations ($N_{ch}$ = constant, $p_{T}$ = constant)\\
$\Sigma_{FB}$ = 1 for independent particle model (IPM), where there is no inter-correlation amongst the produced particles.\\
The values of $\Sigma_{FB} > 1$ has been suggested~\cite{ref1,ref2} to be so due to the effect of Bose-Einstein statistics. Thus, a deviation from the independent emission is clearly observed from the analysis of O-AgBr data sets. Corresponding FRITIOF events, however, do not agree with the experimental results and give $\Sigma_{FB} \sim$ 1 even at the widest $\eta$ window considered. The findings, therefore, suggest the presence of short-range correlations in the data which extends to rather longer range with increasing beam momenta and/or system size.
 In the bottom panel, the ratio plot of model simulated and experimentally measured $\Sigma_{FB}$ values for O-AgBr collisions and for the proton beam induced collisions are displayed. It is evident from these ratio plots that the numerical values of $\Sigma_{FB}$ are underestimated by the model except, in the region $\Delta \eta \le 2$,  where the model overestimates $\Sigma_{FB}$ as compared to the real data set at 14.6A GeV/c. For proton induced collisions, the numerical values of $\Sigma_{FB}$ are underestimated by the model except,  upto $\Delta \eta \le 2$, where the model overestimates the experimentally measured values for p-H and p-AgBr collisions.

 \begin{figure}
\centering
\resizebox{1.0\textwidth}{!}{
\includegraphics{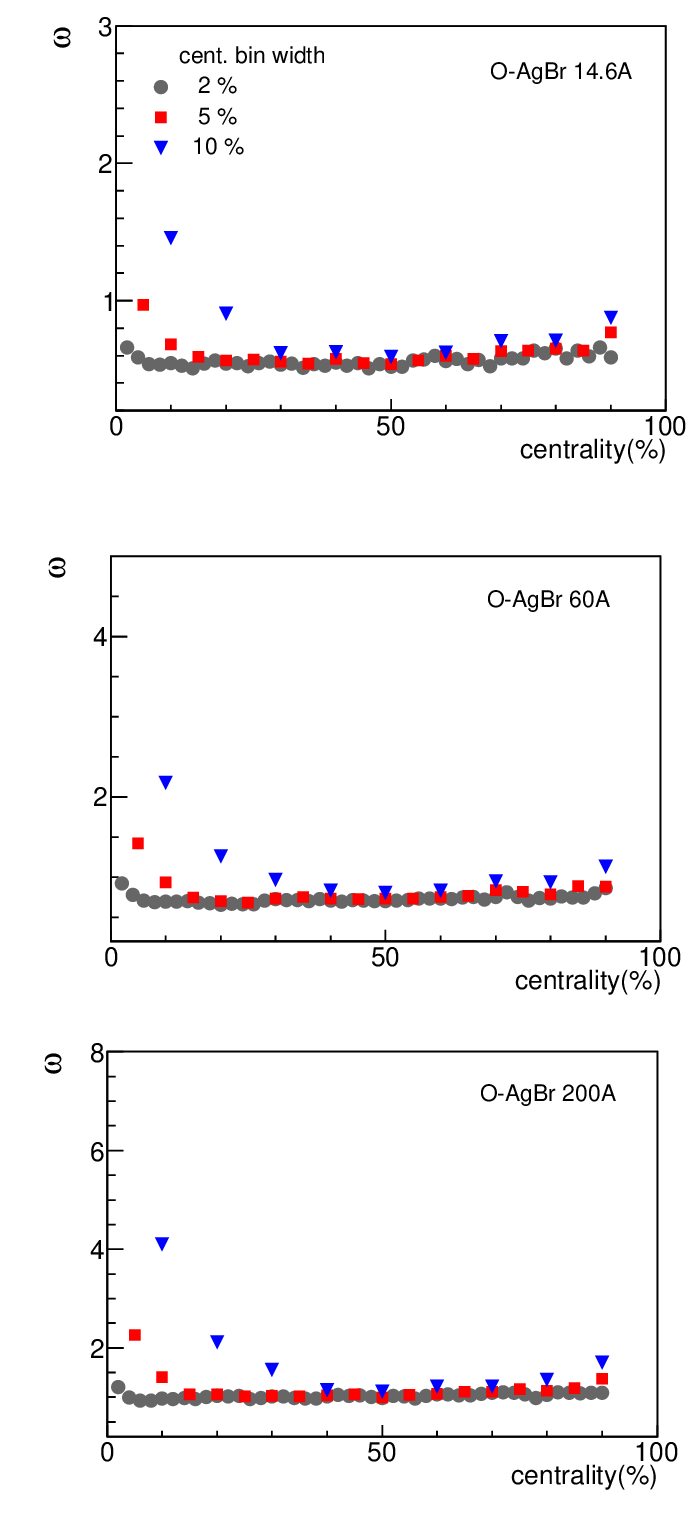}
\includegraphics{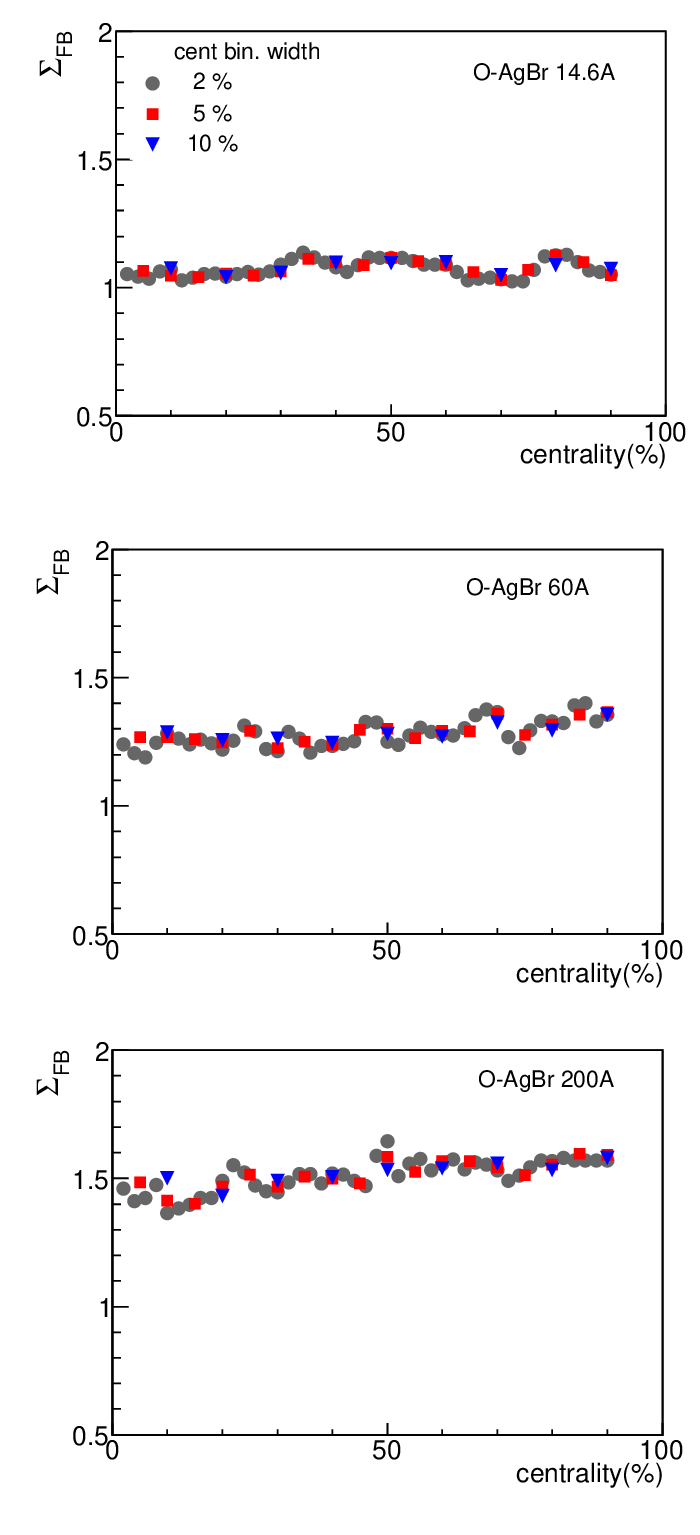} 
 }  
\caption{Centrality bin width dependence of $\omega$(left panel) and $\Sigma_{FB}$ (right panel) for FRITIOF simulated events corresponding to 14.6A, 60A and 200A GeV/c O-AgBr collisions. The error bars shown in the plot are statistical uncertainties only.} 
\label{fig:cent_bin}
 \end{figure}
\subsection{Beam Energy Dependence}
Fluctuation observables are found to exhibit a strong dependence on the projectile beam momentum as evident from Figures~\ref{fig:nch},~\ref{fig:fig5}, and~\ref{fig:Sigma}. The experimental values of the variance $\sigma$, factorial cumulant, $f_2$ and the scaled variance, $\omega$, show a significant growth with increasing beam momenta. It clearly indicates that multiplicity fluctuations are strongly beam momenta dependent and our observation is consistent with the results reported by Alt et. al.~\cite{alt}. It is obvious from O-AgBr collisions that with increasing beam momenta,  $\omega$ values grow significantly due to increased energy density and due to subsequent increased multiplicity density per unit pseudorapidity. Increase in the momenta of oxygen beam results in larger production of charged particles due to higher availability of energy in c.m. frame.  As the multiplicity of charged particle increases, the probability of correlated emission also gets enhanced. The correlated emission observed for charged particles is a clear indicative of the dynamical fluctuations present in the experimental data. For a given number of participating nucleons from projectile and target nuclei, $\Sigma_{FB}$ exhibit a clear dependence on beam momenta. The growing trend of $\Sigma_{FB}$ values found to be greatly pronounced in O-AgBr interactions at beam momenta 200A GeV/c in the entire $\Delta\eta$ range considered in comparison to the corresponding $\Sigma_{FB}$ values at lower beam momenta. At higher beam momenta, the larger energy density of QCD matter is expected to be achieved and the interaction probability of participating nucleons gets increased. The quark and gluon degrees of freedom of the strongly interacting matter becomes dominant, thus, the strength of fluctuations is expected to increase, as is reflected by the anomalous behavior of observables. At lower beam momenta i.e., 14.6A GeV/c, the simulation results on  $\Sigma_{FB}$ as obtained by the Lund Monte Carlo based FRITIOF model code could not mimic the monotonically increasing trend exhibited by O-AgBr interactions and remains more or less constant within statistical errors (as displayed by solid black markers). The findings, thus, clearly display the beam momenta dependence of the fluctuation observables both in real and FRITIOF simulated events (except for beam momenta 14.6A GeV/c).

In order to test whether the correlations and fluctuations exhibited by the observables characterising an event are of dynamical origin, the findings are compared with those obtained from the analysis of the data samples which are free from the dynamical fluctuations. The technique of event mixing gives such a data in which the dynamical correlations amongst the particles are destroyed completely. The mixed event samples corresponding to real and FRITIOF data samples are simulated by adopting the standard procedure of event mixing, described in refs.~\cite{busra,ahmad1,mxd1,mxd2}. Variations of $F_2$, $f_2$ and $\Sigma_{FB}$ with $\Delta\eta$ for the event samples corresponding to the experimental and FRITIOF events are plotted in Figure~\ref{fig:Sigma_mxd}. It may be noted from the figure that $F_2$ values are nearly independent of $\eta$ window width, whereas, these values for the real and FRITIOF data show a clear decreasing trend with increasing $\eta$ window width (Figure~\ref{fig:F2}). $f_2$ values obtained against $\Delta\eta$ for the mixed events are also observed to be smaller than those estimated from the experimental and FRITIOF samples. It has been mentioned earlier in the section 5.4 that $\Sigma_{FB} = 1$ for independent particle production and $\Sigma_{FB} > 1$ would indicate the presence of correlations of some dynamical nature. It is interesting to note in Figure~\ref{fig:Sigma_mxd} that $\Sigma_{FB}$ for the mixed events is $\sim 1$, irrespective of the $\eta$ window width. In the case of real events, however $\Sigma_{FB}$ values are observed to show an increasing trend with $\Delta\eta$ and acquire values $> 1$ for $\Delta\eta > 2$. These observations, thus, indicate that the fluctuations observed with the real data arise due to some dynamical reasons. FRITIOF model also predicts the presence of such fluctuations but the predicted fluctuations are somewhat smaller in magnitude as compared to those observed with the real data.

\subsection{Systematic Uncertainties}
The error bars shown in the figures represent the statistical uncertainties only. As for the systematic uncertainties are concerned, nuclear emulsions, like other detectors, are not free from systematic errors. These unceratinities can, however, be minimized by adopting proper scanning and measurement methods. The method of "along the track" scanning gives reliable event samples because of emulsion's high detection efficiency~\cite{EMU-2,EMU-3,EMU-5}. The scanning is done fast in the forward direction and slow in backward direction. Each pellicle is scanned by two independent observers, so that the biases in detection, counting and measurements can be minimized. This gives scanning efficiency $\gtrsim$ 99\% and hence the systematic errors arising from the scanning are less than 1\%. \\
The presence of backgound tracks, arising due to radioactive contamination may introduce systematic uncertainties~\cite{ebe5,sys52}. Also some of such tracks may appear due to cosmic radiations during exposure, but the chances of such tracks to mix-up with the tracks of genuine events are very very small beacuse the volume occupied by all interaction vertices in emulsion plates is neglegibly small as compared to the total volume of the plates. \\
Production of $e^+e^-$ pair tracks also introduces background contamination. $e^+e^-$ pair tracks due to $\gamma$ conversion do not emanate from the interaction vertex but are produced at certain distance from the interaction vertex after $\gamma$ traverses certain radiation lengths in the medium. To identify the $e^+e^-$ pair tracks from $\gamma$ or $\pi^o$ decay, special care is taken during angular measurements by following the shower tracks upto 100-200 $\mu$m from the interaction vertex. The tracks from $e^+e^-$ pair may easily be identified from their grain densities and are eliminated. \\
Measurements of polar angles would also introduce systematic uncertainties due to fading of tracks~\cite{ebe5}. To prevent the thermal fading, the emulsion pellicles are kept at low temperature and the uncertainties due to fading of tracks becomes too small to effect the analysis for physics studies.
 
 \subsection{Centrality Dependence in the FRITIOF model}
 Since the data collected in emulsion experiments have low tatistics, dividing the data into 10$\%$ or still smaller centrality bins gives very small number of events in each bin, resulting in very high statistical fluctuations. Therefore, in order to study the centrality dependence of fluctuation obervables, FRITIOF events are simulated for O-AgBr collsions at 14.6A, 60A and 200A GeV/c. The number of events in each data set is $3\times10^6$. The centrality of an event is determined by considering the charged particle produced within pseudorapidity $|\eta| < 0.5$. This will reduce the auto-correlation effects~\cite{acoreff} in the estimation of the considered observable in the larger $\eta$ windows. Variations of $\left\langle N_{ch}\right\rangle$ and $\sigma$ with $\Delta \eta$ for the three centrality intervals, 0-5$\%$, 20-30$\%$ and 60-70$\%$ central collsions are shown in Figure~\ref{fig:nch_cent}. It is observed in the figure that with increasing centrality percentile, $\left\langle N_{ch}\right\rangle$ and $\sigma$ decreases at each beam momenta. This is because in central collisions, number of participating nuclei are more which give rise in the multiplicity and in the width of its distribution. Variation of $\omega$ with $\Delta\eta$ for the three central classes are plotted in Figure~\ref{fig:omega_cent}. It may be noted from the figure that for a given $\Delta\eta$, values of $\omega$ are significantly larger for the most central collision as compared to semicentral and peripheral collisions. This  is expected because multiplicity and hence fluctuations in multiplicity would be large in the case of central collisions as compared to semi-central and peripheral collisions. \\

The study of joint fluctuations of two variables in two $\eta$ windows are of special interest as they are connected to the studies of forward-backward (FB) correlations~\cite{ref65,ref66,ref67,ref68}. FB correlations are usually invetigated~\cite{ref67,ref68} in terms of correlation coefficient, which incidently, is not an intensive quantity. FB correlations, if studied in terms of intensive quantities, would permit to examine the intrinsic properties of particle emitting sources because the trivial fluctuations would be suppressed~\cite{ref69}. An attempt has, therefore, been made in the present study to look for FB correlations using the strongly intensive variables $\Sigma_{FB}$, defined earlier. Variations of $\Sigma_{FB}$ with $\Delta\eta$ for 0-5$\%$, 20-30$\%$, 50-60$\%$ central collisions for O-AgBr collisions at 14.6A, 60A and 200A GeV/c for FRITIOF simulated events are plotted in Figure~\ref{fig:sigma_cent}. It may be observed in the figure that $\Sigma_{AB}$ increases with increasing width of the $\eta$ window. It may also be noted in the figure that $\Sigma_{AB}$ shows a faster growth with the $\Delta\eta$ for higher beam energies. Furthermore, It is observed that $\Sigma_{AB}$ values are independent of the centrality class. This indicates the presence of some short range correlations at 60A and 200A GeV/c which extends to rather larger range at higher beam momenta. In order to test whether the observed correlations are due to some dynamical reasons, mixed event sample were obtained from the FRITIOF samples. Results from the analysis of these mixed events are presented in Figure~\ref{fig:sigma_mxd_cent}. It is interesting to note in the figure that $\Sigma_{AB}$ values acquire values $\sim$ 1 irrespective of the beam energy and $\eta$ window widths, which is expected from the independent particle production model which gives $\Sigma_{AB} \sim$ 1. These observations, thus, tend to suggest that the correlation predicted by FRITIOF model are due to some dynamical reasons which are destroyed on mixing the events.\\

Fluctuations in particle multiplicities have significant contributions from statistical and random components. These fluctuations occur due to finite particle multiplicity, centrality slection, effect of re-scattering, etc. Contribution to the fluctuations due to changing impact parameter from even to event, which causes the fluctuations in the number of participating nucleons can be  minimized by proper selection of centrality bin width~\cite{ref70}. Selection of very narrow centrality bin would reduce the geometrical fluctuation to minimum. But a very narrow centrality window would introduce fluctuations due to limited statistics. Hence one has to compromise with somewhat wider centrality windows, e.g., 5$\%$, 10$\%$ of total cross section and then apply the correction to the inherit fluctuations, if required. Centrality dependence of $\omega$ for 2$\%$, 5$\%$ and 10$\%$ centrality intervals for FRITIOF events at various beam momenta are displayed in Figure~\ref{fig:cent_bin}. It is observed that $\omega$ values obtained for 10\% centrality bins are larger as compared to those with 2\% centrality bins. The difference on $\omega$ values is more pronouned for the most central and most peripheral collisions. $\omega$ values for 5\% centrality bins are also observed to be larger than the ones in 2\% centrality bins for the central collisions. This suggests that working with larger centrality bins, $\gtrsim$ 10\% would introduces inherent fluctuations which had to be corrected. It is also interesting to note from Figure~\ref{fig:cent_bin} (right panel), where centrality dependence of $\Sigma_{FB}$ are displayed, that the data points corresponding to various centrality class overlap to form a commom trend. This suggests that for strongly intensive variables, like $\Sigma$ the centrality bin width effect is absent and wider centrality bins may be used for further analysis so as to avoid fluctuations due to limited statistics.  \\

\subsection{Short- and Long-Range Correlations}
 The analysis of forward-backward (F-B) multiplicity correlations has been argued to serve as a potential tool to distinguish between hadron-hadron and parton-parton interactions and  plays a crucial role~\cite{Abelev}. F-B multiplicity correlation can prove to be more informative when looked into short and long-range components. 
Short-range correlations are localized only over a small range of pseudorapidity ($\ <$ 1.0 unit) and arises mainly due to various short-range order effects such as particles produced via cluster decay, resonance decay, or jet correlations~\cite{ahmad}. The particles produced in a single inelastic collision are supposed to exhibit only short-range correlations.   
 If a long-range correlation exists, it should be enhanced in collisions involving nuclei with multiple nucleons. When a system approaches critical point, long-range correlation is expected to exist over a larger pseudo-rapidity range ($\eta$ great than 1 unit). Any possible signature of the string fusion and percolation phenomenon~\cite{FB-1,FB-2,FB-3} in the collisions can be studied by investigating the L-R correlations between multiplicities in two separated rapidity intervals (the forward-backward (F-B) multiplicity correlations) as proposed in ref~\cite{FB-4}.
In this analysis, the data from proton-hydrogen (p-H) collisions is used as a baseline for the measurements of the FB correlation strength that does not include the initial state effects of the colliding nuclei or final state effects from the possible production of a QGP in A-B collisions.  
In Figure~\ref{fig:fig5}, for the lowest considered width of $\eta$ interval, resulting correlation may be considered as coming due to short-range contribution. With increasing width of $\eta$ interval, charged particles from a wider pseudorapidity region are included. Then long-range correlation, if present, would dominate over the short range one as $\sigma$ increases to considerably higher value for A-B collisions. For A-B collisions, the LRC arises mainly due to multiparton interactions. The presence of LRC is predicted for nucleus-nucleus (A-B) collisions in the color glass condensate picture of charged particle production and in the multiple scattering model. FRITIOF simulation data mimic the growing trend but could not exactly reproduce the experimental results quantitatively. 
The short range multiplicity correlation  extending upto larger rage of pseudo rapidity in nucleus-nucleus collisions appear to depend on the system size and also on the beam  as evident from Figure~\ref{fig:Sigma} (real data sets). 
The failure of FRITIOF code to reproduce experimental results quantitatively may be due to the fact that F-B multiplicity correlation as corresponding dynamical input has not been included in the code.

\section{Summary}
The ebe analysis of measured extensive, intensive and strongly intensive quantities has been performed in hadron-hadron (p-H), hadron-nucleus (p-CNO, p-AgBr) and nucleus-nucleus (O-AgBr) collisions with different system size and different beam energies~\cite{ebe5,ebe1,ebe2,ebe3,ebe4,ebe6,ebe7,ebe8,ebe9}. The possible physics explanation for p-A and A-B collisions is interpreted in terms of elementary baseline hadron-hadron (p-H) collisions.
The F-B multiplicity correlations in the terms of $\sigma$, $\omega$ and $\Sigma_{FB}$ with varying $\eta$ window width $\Delta \eta$ for both the real data sets and FRITIOF model predictions, indicate the presence of dynamical fluctuations. For A-B collisions, the magnitude of scaled variance is significantly larger than unity  which  clearly indicates that the multiparticle production in nucleus-nucleus (A-B) collisions is strongly correlated than those observed in hadron-hadron (p-H) collisions and contradicts with the model of independent source emission. The observed features for the second order factorial moments $F_2$ and the factorial cumulant $f_2$ too indicate the presence of dynamical fluctuations in O-AgBr collisions. The correlation strengths (in terms of the intensive variable $\omega$ and strongly intensive quantity $\Sigma$) are found to grow  with the increasing size of pseudorapidity window width  and with the increasing beam momenta for O-AgBr collisions. In case of proton induced collisions, the strength of correlation increases with the increasing size of $\eta$ window width as well as with increasing target size. FRITIOF model predictions are in qualitative agreement with the features observed with the real data, however, the numerical values appear to be rather different. By taking into account the nuclear geometry and  heterogeneity  of  events, FRITIOF model qualitatively reproduces the trend shown by the real data but underestimates the numerical values. The FRITIOF model predicts centrality bin width independent behavior of the strongly intensive quantity $\Sigma_{FB}$. Numerical departure of the FRITIOF model simulation results to replicate the values measured in the experiment may be due to the fact that the FRITIOF model does not well incorporate dynamical parameter inputs responsible for forward-backward multiplicity correlations.

\section{Data Availability}
The data used for the reported findings in this investigation are available with the corresponding author and may be provided on request.

\section{Conflicts of Interest}The authors declare that there are no conflicts of interest.

\end{document}